# Coexistence of polar order and local domain dynamics in ferroelectric SrTi$^{18}$O$_3$

A. Bussmann-Holder, H. Büttner* and A. R. Bishop**

Max-Planck-Institute for Solid State Research, Heisenbergstr.1, D-70569 Stuttgart, Germany

*Lehrstuhl für Theoretische Physik, Universität Bayreuth, D-95540 Bayreuth, Germany

**Los Alamos National Laboratory, Los Alamos, NM 87545, USA

Perovskite oxide ferroelectrics show classical soft mode behaviour typical for the onset of a homogeneous long-range polar state and a displacive phase transition. Besides these long wave length properties, local effects are observed by different probes which reveal that dynamical symmetry breaking already takes place far above the actual instability. It is shown here that displacive mean-field type dynamics can indeed <u>*coexist*</u> with local dynamical symmetry breaking.

Pacs-Index: 77.80.-e, 77.84.-s, 63.20.Kr, 63.20.Ry

Perovskite oxides like, e.g., PbTiO$_3$, BaTiO$_3$, KNbO$_3$, SrTiO$_3$, KTaO$_3$ exhibit a characteristic momentum q=0 transverse optic mode softening which is a precursor of a ferroelectric instability. While in the first three compounds a real instability takes place and a polar state appears at finite temperature, the last two remain stable down to zero temperature because quantum fluctuations suppress the instability and inhibit complete mode softening. These systems are therefore termed incipient ferroelectrics or quantum paraelectrics [1]. This behaviour has attracted considerable interest since quantum effects are usually observed at much lower temperatures. However, early on it has been shown that substitutions in the non-oxide sublattices can easily induce ferroelectricity [2], and simultaneously huge dielectric constants have been observed. Since, in perovskite ferroelectrics with a finite transition temperature $T_c$, isotope effects on $T_c$ have never been observed (in agreement with theoretical predictions [3]), the observation of isotope-induced ferroelectricity in $SrTi^{18}O_3$ [4] has invoked renewed interest in quantum paraelectrics. This observation had, however, been predicted long before its experimental verification since in the quantum limit subtle changes in the effective local potentials occur which are strongly influenced by changes in the sublattice masses [3, 5].

Intense experimental efforts have been made recently in order to understand the microscopic origin of isotope-induced ferroelectricity in $SrTi^{18}O_3$. The results remain, however, quite controversial since local probes like, e.g., NMR and EPR techniques [6, 7] together with birefringence measurements [8] provide evidence that the local structure already differs far above $T_c$ from the high temperature cubic symmetry. Also, the low temperature ferroelectric structure does not have the expected symmetry but seems to be consistent with a mixture of at least two different types of symmetry, suggesting that the ferroelectric transition is not displacive and not of long-range order [9]. The recent observation of perfect transverse momentum q=0 optic mode softening [10] is seemingly in strong contrast with these findings. While the former results support a non mean-field behaviour and non mean-field critical exponents, the latter are in agreement with classical critical exponents as expected at a mean-field level.

We show here that there is in fact no controversy in the interpretation of the different experimental techniques since the time and length scales probed vary considerably. Indeed, the various experimental techniques are complementary.

From the early history of research in ferroelectric perovskites, inelastic neutron scattering experiments have shown that not only a transverse optic mode softens substantially with decreasing temperature to freeze out at $T_c$, but that simultaneously finite q anomalies appear in the related transverse acoustic mode [11, 12, 13]. These anomalies are common to nearly all perovskite ferroelectrics but have never been addressed in detail. What does it mean to have a finite momentum incomplete acoustic mode softening simultaneously appearing with a long wave length optic mode instability? Clearly, the acoustic mode characterizes the elastic stability of the system. Its collapse would indicate a ferroelastic transition where strain/stress plays the role of anorder parameter, whereas the polarisation is the order parameter in a purely ferroelectric system. . The observation of finite momentum anomalies indicates an incipient ferroelastic instability on characteristic length scales, where only finite size domains transform into a new phase. Since softening is never complete in these systems and long wave length coherence is not observed, these anomalies are fingerprints of the formation of _dynamical_ ferroelastic domains which are formed far above $T_c$, and their time and length scales change with temperature. Even though a mode-mode coupling approach has been invoked very early [14], neither a quantitative agreement between theory and experiment could be achieved nor a physical understanding of the microscopic origins. Here we show that mode-mode coupling is indeed the origin, however only so long as polarizability effects are its source. In addition, _we demonstrate that a displacive long wave length soft mode can coexist with local symmetry breaking domains_. We address also the origin of ferroelectricity in isotope substituted SrTi$^{18}$O$_3$ and show that locally distorted dynamical patterns form far above T$_c$ caused by an incipient tendency towards finite size elastic instabilities.

Our results are based on the nonlinear polarizabilty model [15, 16] which relates the local instability of the oxygen 2p$^6$ state [17] with an anharmonic potential in the electron-ion interaction. Opposite to the 2p$^5$ configuration, the O$^{2-}$ one is unstable as a free ion and stabilized in the solid only through the Madelung potential of its surroundings [17]. Changes in the surroundings induce charge redistributions and dynamical covalency effects, which have been shown to be the origin of the ferroelectric instability [18]. We use the polarizability model Hamiltonian given by:

$$H = \frac{1}{2}\sum_n \left[ M_1 \dot{u}_{1n}^2 + m_2 \dot{u}_{2n}^2 + f'(u_{1n+1} - u_{1n})^2 + f(v_{1n} - u_{2n})^2 + f(v_{1n+1} - u_{2n})^2 + g_2 w_{1n}^2 + \frac{1}{2} g_4 w_{1n}^4 \right].$$
(1)

Here, the first terms refer to the kinetic energy of the ionic cores with site *n* dependent displacement coordinates *u*, while the following ones are the potential energies stemming from the interactions between next nearest neighbours (f') and interactions between the shells and the first nearest neighbours (f). The onsite potential in the relative core-shell displacement *w* is anharmonic and consists of an attractive part proportional to g$_2$ and a repulsive fourth order term g$_4$, which together with f' guarantees the stability of the lattice. These onsite terms are a consequence of the oxygen ion 2p$^6$ instability. The mass $M_1$ refers to a cluster representing the BO$_3$ unit in ABO$_3$, whereas $m_2$ represents the A sublattice. The relative core-shell displacement *w* corresponds to an effective polarizability coordinate which measures the long wave length dipole moment. The consequences of equ. 1 are manifold: not only <u>all</u> temperature regimes of the soft mode, including the quantum limit, result self-consistently [16], but also domain wall dynamics [18], local mode formation[19] and anharmonic local displacements patterns are obtained [20]. In addition, as mentioned above, isotope effects have been predicted to vanish in the classical regime, whereas they develop gradually in the quantum regime [3].

Here we concentrate on two results within this model. We address the temperature dependence of the soft long wave length mode and the formation of local regimes with *dynamically* broken symmetry stemming form optic-acoustic mode coupling. Opposite to local effects in the displacements, it is sufficient here to investigate the solutions within the self-consistent phonon approximation (SPA) since these already provide the mode-mode coupling and evidence that dynamical finite size, elastically deformed clusters are formed far above $T_c$. Even though our results apply quite generally to perovskite ferroelectrics, we concentrate in the following on $SrTi^{18}O_3$ since a controversial debate about the origin of ferroelectricity is ongoing for this material. We use the same parameters as in ref. 21 except that $M_1$ has been increased by 10% in order to achieve the experimentally observed value of $T_c$=26K in fully isotope replaced $SrTi^{18}O_3$. All other parameters are the *same* as those used previously [21]. Specifically we note that $SrTi^{16}O_3$, $SrTi^{18}O_3$ and the mass enhanced $SrTi^{18}O_3$ differ only with respect to the mass of $M_1$. Using the SPA, the soft mode is calculated as a function of temperature for $SrTi^{18}O_3$ and compared to the recent experimental results [10] in fig. 1.

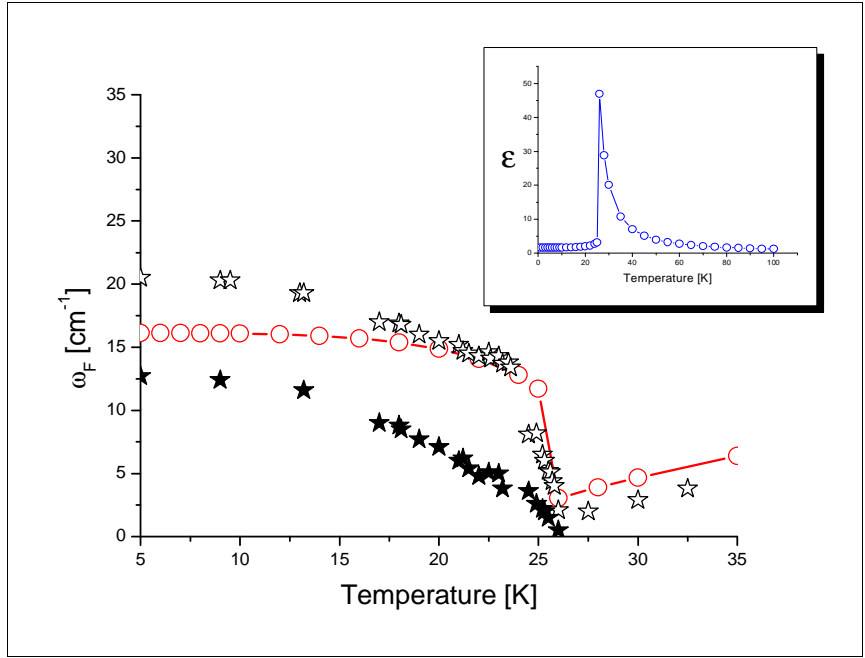

**Figure 1** Comparison of the experimental (black symbols: open stars refer to the $E_{u2}$ mode, full stars to the $E_{u1}$ mode, respectively) [10] and theoretical (red line and open circles) temperature dependence of the soft mode. Above $T_c$ the soft mode is degenerate corresponding to $E_u$ symmetry. The inset shows the calculated temperature dependence of the related dielectric constant $\varepsilon$ as calculated within equ. 1 and the formalism developed in ref. 3.

Even though we are using a pseudo 1D approach (note, that the momentum integrations are over the 3 dimensional Brillouin zone, whereby the existence os a true instability is guaranteed) where the splitting of the soft mode in the low temperature phase can not be accounted for, the agreement with experimental results for the $E_u$ mode is very good in the vicinity of $T_c$; at low temperatures our results correspond to an average of the $E_{u2}$ and $E_{u1}$ soft modes. Even though our results for $\omega_F$ seem to extrapolate to zero at $T_c$, a finite discontinuity remains which evidences the first order character of the transition. The inset to the figure shows the T-dependence of the static inverse dielectric constant calculated within the SPA according to ref. 3. This predicted unusual behaviour in the ferroelectric phase where the dielectric constant is nearly temperature independent, has recently been observed experimentally [22].

Note that the above results only refer to the *long wave length limit* and do not address any *local* effects. We investigated these by examining in detail the acoustic mode dispersion and its evolution with temperature. With decreasing soft mode frequency, the acoustic and optic modes begin coupling at *finite* momentum q, whereby the acoustic mode develops an anomaly in its dispersion reminiscent of a ferroelastic instability. In order to highlight this anomaly, the acoustic mode dispersion is normalized to its value at T=200K where it still follows a harmonic $q^2$ dependence in the long wave length limit. The results of this procedure are shown in figure 2, where the temperature and momentum dependence of the normalized transverse acoustic mode are shown. While for T=180 and 160K the mode still behaves approximately harmonically, an anomaly develops at T=150 K which becomes more pronounced with decreasing temperature. Simultaneously, the critical q value at which it appears shifts to the long wave length limit when approaching $T_c$. However, it never freezes and the q=0 limit is never reached, as should happen when the system forms a homogeneous ground state. These results provide fundamental evidence that dynamical clusters form around 150K with a length scale of approximately 6 lattice constants. With decreasing temperature the clusters increase in size and their dynamics slow down substantially. Around $T_c$ they reach a size of nearly 100 lattice constants and become quasi-static. Below $T_c$ the results are even more striking since the clusters shrink in size and their dynamics become faster again, even though these are still slow as compared to typical phonon frequencies. This behaviour is distinctly different from $SrTi^{16}O_3$ where the cluster size continuously increases with decreasing temperature but always remains substantially smaller than in the present case. However, with increasing mass $M_1$ the cluster size increases as compared to the present case, and a homogeneous ferroelectric state is formed below $T_c$.

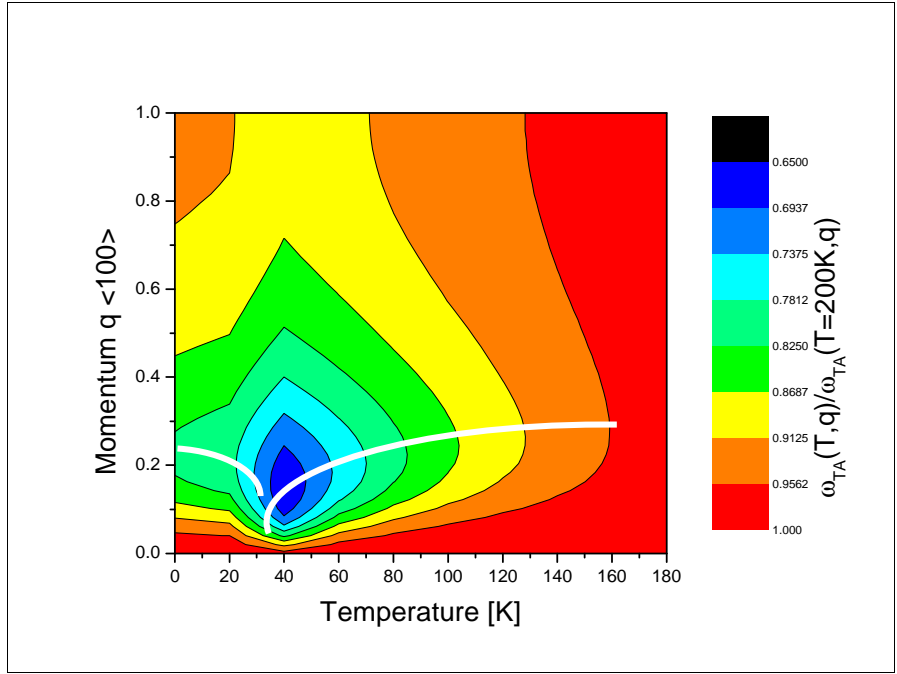

**Figure 2** Momentum and temperature dependence of the normalized transverse acoustic mode. The colour code refers to its relative softening. The white lines refer to the wave vector $q_c$ where the acoustic mode exhibits an anomaly.

This observation evidences very clearly that the ferroelectric state of $SrTi^{18}O_3$ is not conventional. Rather, it is governed by a _coexistence_ of dynamically distorted clusters and ferroelectrically polarized domains so that the ferroelectric state remains _incomplete_. This conclusion supports results from birefringence measurements where the coexistence of different symmetries in the low temperature state has been observed [8, 9]. In addition, NMR experiments in the high and low temperature phases [6, 7] can be understood within this approach since these also report a local symmetry breaking in both phases.

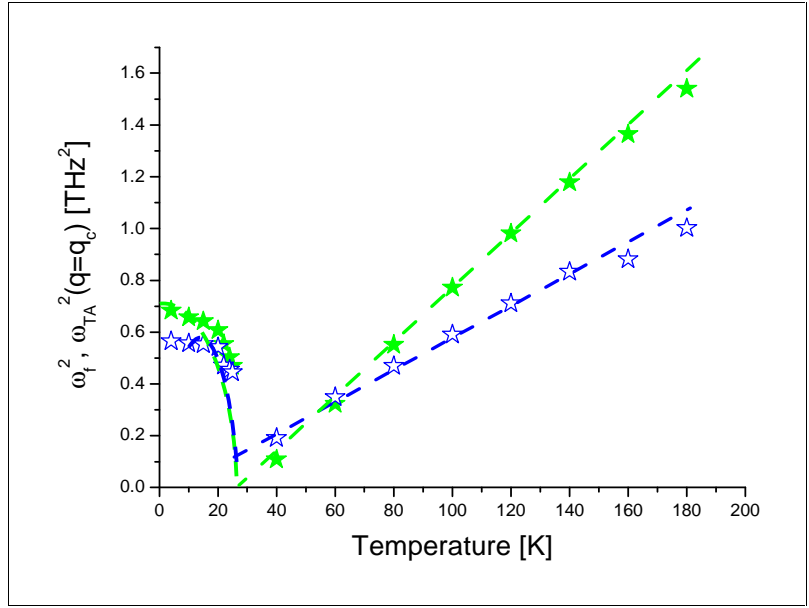

**Figure 3** Calculated temperature dependencies of the squared q=0 ferroelectric mode (green full stars and lines which are a guide to the eye) and the squared acoustic mode frequencies (blue open stars and lines which are a guide to the eye). Here the values of the acoustic mode refer to the critical wave vector $q_c$ where the anomaly appears.

In order to gain more insight into the coexistence of dynamical clusters and the soft mode, we have compared both squared mode frequencies as functions of temperature (fig. 3), where the acoustic mode frequency refers to the one at the anomalous q-value $q=q_c$. In a wide temperature regime, both modes are nearly degenerate and especially below $T_c$ this degeneracy is pronounced. This highlights again that their dynamics coexist and may even compete. This observation is, however, not the same as in $PbTiO_3$ or $KNbO_3$ where the acoustic mode energy is well beyond that of the ferroelectric soft mode and the ferroelectric state is complete. On the other hand, in the quantum paraelectric KTaO$_3$ the dynamics are reversed such that the acoustic mode energy is higher than that of the optic mode and completely governs the dynamical properties [23].

In conclusion, we have shown that the recent finding of a displacive type soft mode in SrTi$^{18}$O$_3$ [10] together with the anomalous low temperature behaviour of the dielectric constant [22] can be consistently explained within a nonlinear polarizability model. The observation of a perfect optic mode softening is, however, not a unique indication that the dynamics of the corresponding system are *exclusively* governed by it. Specifically, the long wave length behaviour does not shed light on local effects which can coexist and even compete. These local dynamics develop already far above the lattice instability and, most importantly, remain active into the ferroelectric phase. Correspondingly, this state is *incomplete* and represents a *coexistence* of two different symmetry states. Our results are consistent with the long wave length results but also with experiments which probe local properties. In terms of a Landau free energy expansion we arrive at a picture of two coupled order parameters, i.e., polarization and strain, where the latter has a temperature dependent length scale.

**Acknowledgements** It is a pleasure to acknowledge stimulating discussions with W. Kleemann, J. F. Scott and R. Blinc. Work at Los Alamos is supported by the USDoE.

**References**

1. K. A. Müller and H. Burkard, Phys. Rev. B **19**, 3593 (1979).

2. J. G. Bednorz and K. A. Müller, Phys. Rev. Lett. **52**, 2289 (1984).

3. A. Bussmann-Holder and H. Büttner, Phys. Rev. B **41**, 9581 (1990).

4. M. Itoh, R. Wang, Y. Inaguma, T. Yamaguchi, Y.-J. Shan and T. Nakamura, Phys. Rev. Lett. **82**, 3540 (1999).

5. A. Bussmann-Holder and A. R. Bishop, Phys. Rev. B **70**, 024104 (2004).

6. R. Blinc, B. Zalar, V. V. Laguta and M. Itoh, Phys. Rev. Lett. **94,** 147601 (2005).

7. V. V. Laguta, R. Blinc, M. Itoh, J. Seliger and B. Zalar, *Phys. Rev. B* **72,** 214117 (2005).

8. J. Dec, W. Kleemann and M. Itoh, Phys. Rev. B **71**, 144113 (2005).

9. L. Zhang, W. Kleemann, R. Wang and M. Itoh, Appl. Phys. Lett. **81**, 3022 (2002).

10. M. Takesada, M. Itoh and T. Yagi, Phys. Rev. Lett. **96**, 227602 (2006).

11. R. A. Cowley, Phys. Rev. **134**, A981 (1964).

12. G. Shirane, J. D. Axe, J. Harada and J. P. Remeika, Phys. Rev. B **2**, 155 (1970).

13. A. Bussmann-Holder, Phys. Rev. B **56**, 10762 (1997).

14. J. D. Axe, J. Harada and G. Shirane, Phys. Rev. B **1**, 1227 (1970).

15. R. Migoni, H. Bilz and D. Bäuerle, Phys. Rev. Lett. **37**, 1155 (1976).

16. H. Bilz, G. Benedek and A. Bussmann-Holder, Phys. Rev. B **35**, 4840 (1987).

17. A. Bussmann, H. Bilz, R. Roenspiess and K.-H. Schwarz, Ferroelectrics **25**, 343 (1980).

18. G. Benedek, H. Bilz and A. Bussmann-Holder, Phys. Rev. B **36**, 630 (1987).

19. A. Bussmann-Holder and A. R. Bishop, Phys. Rev. B **70**, 184303 (2004).

20. A. Bussmann-Holder, A. R. Bishop and G. Benedek, Phys. Rev. B **53**, 11521 (1996).

21. A. Bussmann-Holder, H. Büttner and A. R. Bishop, J. Phys.: Cond. Mat. **12**, L115 (2000).

22. C. Filipič and A. Levstik, Phys. Rev. B **73**, 092104 (2006).

23. A. Bussmann-Holder, A. R. Bishop and H. Büttner, unpublished.